\documentclass[twocolumn,showpacs,preprintnumbers,amsmath,amssymb,pra]{revtex4}
\usepackage{graphicx}
\usepackage{dcolumn}
\usepackage{bm}
\usepackage{times}
\usepackage[colorlinks,citecolor=blue,linkcolor=red]{hyperref}
\usepackage{color}
\usepackage{comment}

\begin{document}

\title{Tuning nonequilibrium heat current and two-photon statistics via composite qubit-resonator interaction}

\author{Zhe-Huan Chen$^{1}$}
\author{Han-Xin Che$^{1}$}
\author{Zhe-Kai Chen$^{1}$}
\author{Chen Wang$^{1,}$}\email{wangchenyifang@gmail.com; wangchen@zjnu.cn}
\author{Jie Ren$^{2,}$}\email{Xonics@tongji.edu.cn}
\address{
$^{1}$Department of Physics, Zhejiang Normal University, Jinhua 321004, Zhejiang, P. R. China\\
$^{2}$Center for Phononics and Thermal Energy Science, China-EU Joint Center for Nanophononics, \\
Shanghai Key Laboratory of Special Artificial Microstructure Materials and Technology,  \\
School of Physics Sciences and Engineering, Tongji University, Shanghai 200092, China
}
\date{\today}

\begin{abstract}
Quantum thermal transport and two-photon statistics serve as two representative nonequilibrium features in circuit quantum electrodynamics systems.
Here, we investigate quantum heat flow and two-photon correlation function at steady-state in a composite qubit-resonator model,
where one qubit shows both transverse and longitudinal couplings to a single-mode optical resonator.
With weak qubit-resonator interaction,
we unravel two microscopic transport pictures, i.e., cotunneling and cyclic heat exchange processes,
corresponding to transverse and longitudinal couplings respectively.
At strong qubit-resonator coupling,
the heat current exhibits nonmonotonic behavior by increasing qubit-resonator coupling strength,
which tightly relies on the scattering processes between the qubit and corresponding thermal bath.
Furthermore, the longitudinal coupling is preferred to enhance heat current in strong qubit-resonator coupling regime.
For two-photon correlation function, it exhibits an antibunching-to-bunching transition,
which is mainly dominated by the modulation of energy gap between the first and second excited eigenstates.
Our results are expected to deepen the understanding of nonequilibrium thermal transport and nonclassical photon radiation
based on the circuit quantum electrodynamics platform.
\end{abstract}


\maketitle

\section{Introduction}
Deep understanding and efficient characterization of nonequilibrium energy transport via quantum light-matter interaction constitutes an active frontier for quantum optics and quantum transport~\cite{gchen2005book,hxu2016nature,plodahl2017nature,aronzani2018,dwwang2019np}.
The heat flow is considered as one generic feature of quantum thermal transport, which is bounded by the second law of thermodynamics.
Under thermodynamic bias (e.g., voltage and temperature bias),
the heat current is driven directionally from the hot source to the cold drain.
However, the direction of the current can be reversed against the thermodynamic bias, e.g.,
by quantum correlations~\cite{kmicadei2019nc} and geometric-phase-induced pump~\cite{jren2010prl,zwang2021fop}.

Due to the dramatic advance of quantum circuit technology, the circuit quantum electrodynamics (cQED) systems emerge as one representative platform to realize quantum light-matter interaction~\cite{gkurizki2015pnas,ab2020np,aaclerk2020np,ab2021rmp}.
The cQED systems are traditionally described by the seminal quantum Rabi model (QRM)~\cite{rabi1,rabi2,braak2011prl,qhchen2012pra}, i.e., one two-level qubit transversely interacting with a single-model photon resonator,
which is able to describe ultrastrong qubit-resonator coupling~\cite{pfdiaz2019rmp,afkockum2019nrp,lb2020aqt}.
QRM has been extensively investigated in finite-component quantum phase transition~\cite{mjhwang2015prl,mxliu2017prl,yyzhang2021prl,mlcai2021nc},
quantum nonlinear optics~\cite{lgarziano2015pra,lgarziano2016prl,xwang2017pra,vmacri2021arxiv}, and quantum thermodynamics~\cite{jhwang2012pre,faltintas2015pra,sseah2018pre}.
In particular for two-photon statistics, Ridolfo \emph{et al.}~\cite{aridolfo2012prl,aridolfo2013prl,rstassi2013prl,lgarziano2017acs} proposed a modified definition of two-photon correlation function to properly characterize
photon nonclassicality at strong qubit-photon coupling.
Meanwhile, an alternative scheme, i.e., longitudinal coupling between the qubit and the resonator,
can also be realized based on the superconducting circuit engineering~\cite{tniemczyk2010np,fyoshihara2017np},
which has pronounced consequences for nonclassical-photon-state generation~\cite{yjzhao2015pra,xinwang2016pra},
scalable circuit design~\cite{pmb2015prb,sricher2016prb}, and fast nondemolition qubit readout~\cite{ndidier2015prl,algrimsmo2019prb}.


Recently, quantum thermal transport in cQED systems has attracted increasing attention, which leads to a flurry of valuable works~\cite{aronzani2018np,jsenior2020cp,mmajland2020prb,jppekola2021arxiv,ylu2021arxiv}.
Particularly, Pekola \emph{et al.}~\cite{aronzani2018np} experimentally detected heat flow in a hybrid quantum system comprising one transmon-qubit and two microwave resonators,
of which the resonators are individually coupled to two metallic resistors, respectively.
Consequently, the typical thermal functionalities were realized, e.g., heat valve~\cite{aronzani2018np},
thermal diode~\cite{jsenior2020cp}, and thermal transistor~\cite{mmajland2020prb}.
Moreover, Maguire \emph{et al.}~\cite{jsmith2014pra,jsmith2016jcp,hmaguire2019prl} theoretically investigated Franco-Condon physics in noncommutative QRM via the reaction coordinate mapping approach.
Yamamoto \emph{et al.}~\cite{tyamamoto2021jpcm} unravelled nontrivial two-peak feature of thermal conductance in QRM at linear response limit.
Wang \emph{et al.}~\cite{cwang2021cpl,cwang2021cpb} analyzed nonmonotonic behavior of the heat current in a longitudinally-coupled qubit-resonator model.
However,  nonequilibrium heat flow and the microscopic picture in dissipative QRM at finite temperature bias currently is lack of exploration,
which is crucial to deepen the understanding of nonequilibrium heat transport based on the cQED platform.
Furthermore, though both the transverse and longitudinal couplings, between the qubit and the optical resonator can be flexibly modulated~\cite{tniemczyk2010np,fyoshihara2017np,lgarziano2016prl,xinwang2016pra,nlambert2018prb},
the influence of  composite qubit-resonator interaction in the heat current and two-photon statistics has not yet been reported.

In this paper, we applied dressed master equation (DME) to investigate the effect of composite qubit-resonator interaction on quantum thermal transport.
At weak qubit-resonator coupling, it is found
the heat current exhibits monotonic enhancement by increasing thermal bath temperature bias in the dissipative QRM,
which stems from cotunneling transport processes.
In contrast, the current is changed to show nonmonotonic feature with longitudinal qubit-resonator coupling,
which is dominated by cyclic heat exchange transitions.
These two distinct microscopic processes are crucial to unravel the physics pictures of quantum thermal transport
in dissipative cQED systems.
At strong qubit-resonator coupling with different composite angles,
the current generally shows nomonotonic behavior,
{which tightly relies on the scattering processes between the qubit and bosons in the corresponding thermal bath.}
Furthermore, the optimal composite angle to generate the maximal heat current, gradually changes from the transverse coupling type to the longitudinal counterpart with increase of qubit-resonator coupling strength.
We also study steady-state two-photon correlation function at strong qubit-resonator coupling.
It is found that by tuning up the composite angle, an antibunching-to-bunching transition is significantly exhibited, which mainly originates from
increment and reduction of the energy gap between the first and second excited states.
In particular, a giant bunching signature of photons  is unravelled at moderate composite angle.

The rest of this paper is organized as follows:
In section II, we present the composite qubit-resonator model, derive the dressed master equation, and obtain the expression of the steady-state heat current.
In section III, we investigate the effect of the composite angle on the steady-state heat current and two-photon correlation function.
The microscopic pictures of these behaviors are also discussed.
Finally, we give a summary in section IV.
\begin{figure}[tbp]
\begin{center}
\includegraphics[scale=0.5]{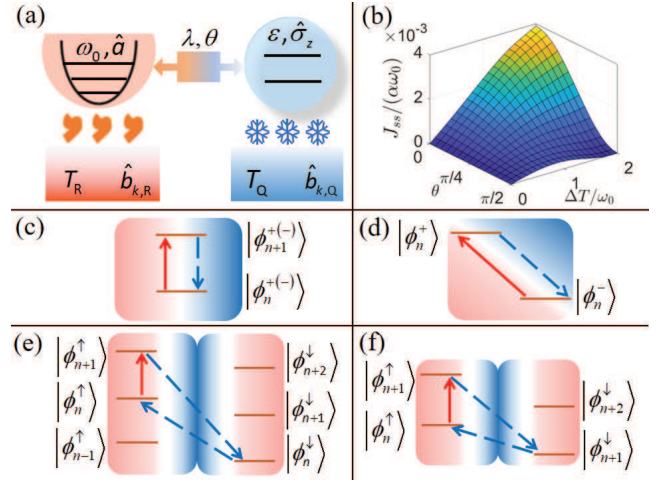}
\end{center}
\caption{(Color online) (a) The schematic description of a composite qubit-resonator model.
The red half-circle (top-left) is the optical resonator, where $\hat{a}$ annihilates one photon with the frequency $\omega_0$.
The blue circle (top-right) represents the qubit, which is characterized by the Pauli operator $\hat{\sigma}_z$ and the splitting energy $\varepsilon$.
The double-arrowed rectangle shows the composite qubit-resonator interaction
with the interaction strength $\lambda$ and composite angle $\theta$.
The red (bottom-left) and blue (bottom-right) rectangles denote two thermal baths, characterized by  temperatures $T_{\textrm{R}}$ and $T_{\textrm{Q}}$, and bosonic annihilators $\hat{b}_{k,\textrm{R}}$ and $\hat{b}_{k,\textrm{Q}}$.
(b) Steady-state heat current $J_{ss}/(\alpha\omega_0)$ modulated by temperature bias ${\Delta}T/\omega_0$ and the composite angle $\theta$,
with weak qubit-resonator interaction strength $\lambda=0.01\omega_0$.
Other parameters are given by $\varepsilon=1.5\omega_0$, $\alpha=0.001$,  $\omega_c=10\omega_0$,
$T_{\textrm{R}}=\omega_0+{\Delta}T/2$, and $T_{\textrm{Q}}=\omega_0-{\Delta}T/2$.
(c) and (d) show cotunneling processes for first terms of the components $I_{x,1}$ and $I_{x,2}$ in Eqs.~(\ref{ix1}-\ref{ix2}), respectively.
(e) and (f) describe cyclic energy exchange processes for first terms of current components $I_{z,1}$ and $I_{z,2}$ in Eqs.~(\ref{iz1}-\ref{iz2}), respectively.
The red solid (blue dashed) arrowed lines denote transitions between two eigenstates  assisted by the $\textrm{R}$-th~($\textrm{Q}$-th) thermal bath.
}~\label{fig1}
\end{figure}

\section{Model and method}
\subsection{Composite qubit-resonator model}
We study the dissipative qubit-resonator hybrid model in Fig.~\ref{fig1}(a), where one two-level qubit shows both longitudinal and transverse couplings to a single-mode resonator, each individually interacting with the corresponding thermal baths.
The Hamiltonian is described as ($\hbar=1$ and $k_B=1$)
\begin{eqnarray}~\label{htot}
\hat{H}=\hat{H}^\theta_{\textrm{S}}+\sum_{\mu=\textrm{Q},\textrm{R}}(\hat{H}^\mu_\textrm{B}+\hat{V}_\mu).
\end{eqnarray}
Specifically, the composite  qubit-resonator system is expressed as
\begin{eqnarray}~\label{hs}
\hat{H^\theta_{\textrm{S}}}=\frac{\varepsilon}{2}\hat{\sigma}_z +\omega_0 \hat{a}^\dagger \hat{a}+\lambda(\cos\theta\hat{\sigma}_x+\sin\theta\hat{\sigma}_z)(\hat{a}^\dagger +\hat{a}),
\end{eqnarray}
where $\hat{a}^\dagger(\hat{a})$ is the creation (annihilation) operator of one photon in the resonator with the frequency $\omega_0$,
$\varepsilon$ is the splitting energy of the two-level qubit,
$\hat{\sigma}_\alpha~(\alpha=x,y,z)$ is the Pauli operators of the qubit under the qubit basis $\{\uparrow, \downarrow\}$,
with $\hat{\sigma}_z|\uparrow{\rangle}=|\uparrow{\rangle}$ and $\hat{\sigma}_z|\downarrow{\rangle}=-|\downarrow{\rangle}$,
$\lambda$ is the qubit-resonator coupling strength,
and $\theta$ is the angle to tune the composite qubit-resonator interaction.
$\hat{H}^\mu_\textrm{B}=\sum_k\omega_{k,\mu}\hat{b}^\dag_{k,\mu}\hat{b}_{k,\mu}$ describes the $\mu$-th bosonic thermal bath,
where $\hat{b}^\dag_{k,\mu}~(\hat{b}_{k,\mu})$ creates (annihilates) one boson with the frequency $\omega_{k,\mu}$ and momentum $k$.
The system-bath interactions are described as
\begin{subequations}
\begin{align}
\hat{V}_\textrm{R}=&(\hat{a}^\dagger+\hat{a})\sum_k g_{k,\textrm{R}}(\hat{b}^\dagger_{k,\textrm{R}}+\hat{b}_{k,\textrm{R}}),~\label{vr}\\
\hat{V}_\textrm{Q}=&\hat{\sigma}_x\sum_k g_{k,\textrm{Q}}(\hat{b}_{k,\textrm{Q}}^\dagger+\hat{b}_{k,\textrm{Q}}),~\label{vq}
\end{align}
\end{subequations}
with $g_{k,\mu}$ the coupling strength.
The $\mu$-th thermal bath is characterized by the spectral function $\gamma_\mu(\omega)=2\pi\sum_k|g_{k,\mu}|^2\delta(\omega-\omega_{k,\mu})$.
In this work, we specify the spectral function as the Ohmic case, i.e., $\gamma_\mu(\omega)=\pi\alpha\omega\exp(-|\omega|/\omega_c)$,
with $\alpha$ the dissipation strength and $\omega_c$ the cutoff frequency.

Generally, it is difficult to analytically find the eigensoultion of the qubit-resonator hybrid system at Eq.~(\ref{hs}).
However, in the limiting case we may obtain the analytical expression.
Specifically, at $\theta=0$ the composite system  with only transverse coupling is simplified as
\begin{eqnarray}
\hat{H}^{0}_{\textrm{S}}=\frac{\varepsilon}{2}\hat{\sigma}_z +\omega_0 \hat{a}^\dagger \hat{a}+\lambda\hat{\sigma}_x(\hat{a}^\dagger +\hat{a}),
\end{eqnarray}
which is the seminal quantum Rabi model~\cite{rabi1,rabi2}.
Hence, the eigenvalues can be mapped to the roots of transcendental G-function by applying the Bargmann algebra and extended coherent states approaches~\cite{braak2011prl,qhchen2012pra}, respectively.
In  weak qubit-resonator coupling regime, the quantum Rabi model is reduced to the Jaynes-Cummings model
$\hat{H}_{\textrm{JC}}={\varepsilon}\hat{\sigma}_z/2 +\omega_0 \hat{a}^\dagger \hat{a}
+\lambda(\hat{a}^\dagger\hat{\sigma}_- +\hat{a}\hat{\sigma}_+)$,
which is dominated by the rotating-wave-terms.
Consequently, the eigenvalues are given by
\begin{eqnarray}
E_{n,\pm}=(n+1/2)\omega_0{\pm}\sqrt{(\varepsilon-\omega_0)^2/4+\lambda^2(n+1)},
\end{eqnarray}
and the corresponding eigenvectors are
\begin{subequations}
\begin{align}
|\phi^+_n{\rangle}=&\cos\frac{\theta_n}{2}|n,\uparrow{\rangle}+\sin\frac{\theta_n}{2}|n+1,\downarrow{\rangle},~\label{vecxp}\\
|\phi^-_n{\rangle}=&-\sin\frac{\theta_n}{2}|n,\uparrow{\rangle}+\cos\frac{\theta_n}{2}|n+1,\downarrow{\rangle},~\label{vecxm}
\end{align}
\end{subequations}
with $\tan\theta_n=2\lambda\sqrt{n+1}/(\varepsilon-\omega_0)$.

While at $\theta=\pi/2$ the qubit is longitudinally coupled to the resonator~\cite{yjzhao2015pra,sricher2016prb,xinwang2016pra,pmb2015prb,nlambert2018prb,cwang2021cpl}, with the Hamiltonian
\begin{eqnarray}
\hat{H}^{\pi/2}_{\textrm{S}}=\frac{\varepsilon}{2}\hat{\sigma}_z +\omega_0 \hat{a}^\dagger \hat{a}+\lambda\hat{\sigma}_z(\hat{a}^\dagger +\hat{a}).
\end{eqnarray}
Accordingly, the eigenvalues are expressed as
\begin{subequations}
\begin{align}
E_{n,\uparrow}=&\omega_0n+\varepsilon/2-\lambda^2/\omega_0,\\
E_{n,\downarrow}=&\omega_0n-\varepsilon/2-\lambda^2/\omega_0,
\end{align}
\end{subequations}
and the eigenvectors are described by the extended coherent boson states
\begin{subequations}
\begin{align}
|\phi^{\uparrow}_n{\rangle}=&\exp\Big[\frac{\lambda}{\omega_0}(\hat{a}-\hat{a}^\dag)\Big]\frac{(\hat{a}^\dag)^n}{\sqrt{n!}}|0{\rangle}_a{\otimes}|\uparrow{\rangle},~\label{veczp}\\
|\phi^{\downarrow}_n{\rangle}=&\exp\Big[-\frac{\lambda}{\omega_0}(\hat{a}-\hat{a}^\dag)\Big]\frac{(\hat{a}^\dag)^n}{\sqrt{n!}}|0{\rangle}_a{\otimes}|\downarrow{\rangle},~\label{veczm}
\end{align}
\end{subequations}
with the vacuum state of the resonator $\hat{a}|0{\rangle}_a=0$.

We should note that the dissipative composite qubit-resonator model is more than a toy model.
It can be experimentally realized in the superconducting quantum circuit platforms~\cite{aronzani2018np,jsenior2020cp,jppekola2021arxiv,mmajland2020prb}, where the transmon qubit is able to show both the longitudinal and transverse interactions with the microwave resonator,
and the bosonic thermal bath is simulated by the LC circuit coupled to a metallic resistor.
Hence, the heat energy naturally flows from the hot source to the cold drain under the temperature bias.

\subsection{Quantum master equation}
We consider weak interactions between the composite hybrid system and bosonic thermal baths.
We focus on the steady-state properties of the composite qubit-resonator system, where the off-diagonal elements
of the reduced system density operator in the eigenbasis of $\hat{H}_{\textrm{S}}$ become negligible.
Thus, $\hat{V}_{\textrm{R}}$ and $\hat{V}_{\textrm{Q}}$ at Eqs.~(\ref{vr}-\ref{vq}) can be properly perturbed.
Under the Born approximation, the total density operator can be separated as
$\hat{\rho}_{\textrm{tot}}(t){\approx}\hat{\rho}_{\textrm{S}}(t)
{\otimes}\hat{\rho}_{\textrm{B},\textrm{R}}{\otimes}\hat{\rho}_{\textrm{B},\textrm{Q}}$,
where $\hat{\rho}_{\textrm{S}}(t)$ is the reduced hybrid system density operator,
and $\hat{\rho}_{\textrm{B},\mu}
=\exp(-\hat{H}^{\mu}_\textrm{B}/k_\textrm{B}T_u)/\textrm{Tr}_\textrm{B}\{\exp(-\hat{H}^{\mu}_\textrm{B}/k_\textrm{B}T_u)\}~(\mu=\textrm{R},\textrm{Q})$ is the density operator of the $\mu$-th thermal bath, with $k_\textrm{B}$ the Planck constant
and $T_\mu$ is the temperature of the $\mu$-th bath.
Then, by further including the Markov approximation, we obtain the quantum dressed master equation as~\cite{fbeaudoin2011pra,asettineri2018pra,lb2020aqt}
\begin{eqnarray}~\label{dme}
\frac{d}{dt}\hat{\rho}_{\textrm{S}}(t)&=&i[\hat{\rho}_{\textrm{S}}(t),\hat{H}^\theta_{\textrm{S}}]
+\sum_{n,m,\mu}\{\Gamma^+_{\mu}(E_{nm})\hat{\mathcal{L}}_{nm}[\hat{\rho}_{\textrm{S}}(t)]\nonumber\\
&&+\Gamma^-_{\mu}(E_{nm})\hat{\mathcal{L}}_{mn}[\hat{\rho}_{\textrm{S}}(t)]\},
\end{eqnarray}
where the dissipator is given by
$\hat{\mathcal{L}}_{nm}[\hat{\rho}_{\textrm{S}}(t)]
=|\phi_n{\rangle}{\langle}\phi_m|\hat{\rho}_{\textrm{S}}(t)|\phi_m{\rangle}{\langle}\phi_n|
-(|\phi_m{\rangle}{\langle}\phi_m|\hat{\rho}_{\textrm{S}}(t)+\hat{\rho}_{\textrm{S}}(t)|\phi_m{\rangle}{\langle}\phi_m|)/2$,
with $\hat{H}^\theta_{\textrm{S}}|\phi_n{\rangle}=E_n|\phi_n{\rangle}$,
and the corresponding transitions rates are given by
\begin{subequations}
\begin{align}
\Gamma^+_{\mu}(E_{n,m})=&\gamma_\mu(E_{n,m})n_{\mu}(E_{n,m})|{\langle}\phi_n|\hat{A}_\mu|\phi_m{\rangle}|^2,~\label{ratep}\\
\Gamma^-_{\mu}(E_{n,m})=&\gamma_\mu(E_{nm})[1+n_{\mu}(E_{n,m})]|{\langle}\phi_n|\hat{A}_\mu|\phi_m{\rangle}|^2,~\label{ratem}
\end{align}
\end{subequations}
with $n_{\mu}(E_{n,m})=1/[\exp(E_{n,m}/T_u)-1]$ the Bose-Einstein distribution function,
$\hat{A}_{\textrm{R}}=\hat{a}^\dag+\hat{a}$,
$\hat{A}_{\textrm{Q}}=\hat{\sigma}_x$,
$E_{n,m}=E_n-E_m$ the energy gap between two eigenstates $|\phi_n{\rangle}$ and $|\phi_m{\rangle}$ of $\hat{H}_{\textrm{S}}$.
The rate $\Gamma^{+(-)}_{\mu}(E_{n,m})$ describes the energy exchange process that the composite qubit-resonator system is excited (relaxed) from the eigenstate $|\phi_m{\rangle}$ to $|\phi_n{\rangle}$ by absorbing (releasing) one photon with the energy $E_n-E_m$ from (into) the $\mu$-th thermal bath.

Therefore, after long time evolution, i.e. $d\hat{\rho}_{\textrm{S}}(t)/dt=0$, we can obtain the steady-steady population distribution
$P_{n}$ with $P_{n}={\langle}\phi_n|\hat{\rho}_{\textrm{S}}(t\rightarrow\infty)|\phi_n{\rangle}$.
Moreover, from the dressed master equation Eq.~(\ref{dme})  we obtain the steady-state heat current into the $\textrm{Q}$-th bath (see part A of the Appendix)
\begin{eqnarray}
J_{ss}=\sum_{E_n>E_{n^\prime}}E_{n,n^\prime}[\Gamma^-_{\textrm{Q}}(E_{n,n^\prime})P_{n}-\Gamma^+_{\textrm{Q}}(E_{n,n^\prime})P_{n^{\prime}}],
\end{eqnarray}
where the energy gap is $E_{n,n^\prime}=E_n-E_{n^\prime}$.

\section{Results and discussions}
\subsection{Steady state heat current}

\subsubsection{Weak qubit-resonator interaction}
We first investigate the steady-state behavior of  heat current at weak qubit-resonator coupling in Fig.~\ref{fig1}(b),
which is both modulated by the temperature bias ${\Delta}T=T_{\textrm{R}}-T_{\textrm{Q}}$ and composite angle $\theta$.
It is found that for small $\theta$, the heat current exhibits monotonic increase by increasing the temperature bias ${\Delta}T$,
particularly in the limit of $\theta=0$, i.e., the dissipative quantum Rabi model.
However, in the large $\theta$ regime the heat current is changed to shown nonmonotonic behavior,
i.e., the current is first enhanced and later suppressed with the increase of ${\Delta}T$,
which identifies the signature of the negative differential thermal conductance (NDTC)~\cite{bli2006apl,bli2012rmp,dhhe2009prb,dhhe2010pre,hkchan2014pre}.
The appearance of NDTC is consistent with previous works~\cite{cwang2021cpl,cwang2021cpb}.
It needs to note that though not shown here,
the similar result can also be  found at resonance ($\varepsilon=\omega_0$).
Hence, we conclude that the composite qubit-resonator interaction strongly affects the steady-state heat current.

Then, we try to explore microscopic processes of heat transport with weak qubit-resonator coupling.
We admit that to analytically find the microscopic mechanism with arbitrary composite angle $\theta$ is quite difficult.
Here, we focus on two limits, i.e., $\theta=0$ and $\theta=\pi/2$, to unravel the representative physical pictures of the heat current.
For $\theta=0$, under the eigenbasis $\{|\phi^{\pm}_n{\rangle}\}$ of QRM in Eqs.~(\ref{vecxp}-\ref{vecxm}), the leading order of steady-state heat current at finite energy bias regime [e.g., $(\varepsilon-\omega_0){\gg}\lambda$] can be analytically expressed as (see the detail in part B of the Appendix)
\begin{eqnarray}~\label{jxss}
J^{x}_{ss}&{\approx}&\frac{\lambda^2}{(\varepsilon-\omega_0)^2}(\omega_0I_{x,1}+{\varepsilon}I_{x,2}),
\end{eqnarray}
where two components are specified as
\begin{subequations}
\begin{align}
I_{x,1}=&\gamma_{\textrm{Q}}(\omega_0)\{n_{\textrm{R}}(\omega_0)[1+n_{\textrm{Q}}(\omega_0)]-[1+n_{\textrm{R}}(\omega_0)]n_{\textrm{Q}}(\omega_0)\},~\label{ix1}\\
I_{x,2}=&\frac{\gamma_{\textrm{R}}(\varepsilon)}{2n_{\textrm{Q}}(\varepsilon)+1}
\{n_{\textrm{R}}(\varepsilon)[1+n_{\textrm{Q}}(\varepsilon)]
-{[1+n_{\textrm{R}}(\varepsilon)]}n_{\textrm{Q}}(\varepsilon)\}.~\label{ix2}
\end{align}
\end{subequations}
Both $I_{x,1}$ and $I_{x,2}$ are dominated by  cotunneling processes.
Specifically, $I_{x,1}$ describes the process such that as the state $|\phi^{\eta}_n{\rangle}$ $(|\phi^{\eta}_{n+1}{\rangle})$ is excited (relaxed) to $|\phi^{\eta}_{n+1}{\rangle}$ $(|\phi^{\eta}_n{\rangle})$ by absorbing (emitting)
energy $\omega_0$ from (into) the $\textrm{R}$-th reservoir, the transition $|\phi^{\eta}_{n+1}{\rangle}{\rightarrow}|\phi^{\eta}_n{\rangle}$
$(|\phi^{\eta}_{n}{\rangle}{\rightarrow}|\phi^{\eta}_{n+1}{\rangle})$
simultaneously occurs by emitting (absorbing) $\omega_0$ into (from) the $\textrm{Q}$-th reservoir with $\eta=\pm$,
which is also shown in Fig.~\ref{fig1}(c).
While $I_{x,2}$ shows other typical cotunneling processes, exemplified in Fig.~\ref{fig1}(d),
that the excitation (relaxation) transition
$|\phi^{-(+)}_{n}{\rangle}{\rightarrow}|\phi^{+(-)}_{n}{\rangle}$ by absorbing (emitting) energy $\varepsilon$ from (into) the $\textrm{R}$-th reservoir is accompanied by the dual transition $|\phi^{+(-)}_{n}{\rangle}{\rightarrow}|\phi^{-(+)}_{n}{\rangle}$.
It is interesting to find that the directional cotunneling transport from the $\textrm{R}$-th reservoir to $\textrm{Q}$-th one, i.e. described by first terms of $I_{x,1}$ and $I_{x,2}$, are monotonically enhanced with increase of the temperature bias,
which mainly contribute to $J^x_{ss}$ at finite temperature bias.
In contrast, the opposite transitions (from the $\textrm{Q}$-th reservoir to $\textrm{R}$-th one) is dramatically suppressed.
Finally, the current in Eq.~(\ref{jxss}) becomes significant at large temperature bias $(T_{\textrm{R}}{\approx}2\omega_0,~T_{\textrm{Q}}{\approx}0)$,
which is specified as
$J^x_{ss}{\approx}[{\lambda}/{(\varepsilon-\omega_0)}]^2
[\omega_0\gamma_{\textrm{Q}}(\omega_0)n_{\textrm{R}}(\omega_0)
+{\varepsilon}\gamma_{\textrm{R}}(\varepsilon)n_{\textrm{R}}(\varepsilon)]$.

While for $\theta=\pi/2$, it is found that the transition coefficient ${\langle}\phi^{\sigma}_{n}|\hat{\sigma}_x|\phi^{\overline{\sigma}}_{n^{\prime}}{\rangle}$ in Eqs.~(\ref{ratep}-\ref{ratem}) under the coherent-state basis $\{|\phi^{\uparrow(\downarrow)}_n{\rangle}\}$ is approximated as
${\langle}\phi^{\uparrow}_{n}|\hat{\sigma}_x|\phi^{\downarrow}_{n^{\prime}}{\rangle}
{\approx}(-1)^n\Big[\delta_{n,n^\prime}+({2\lambda}/{\omega_0})(\sqrt{n+1}\delta_{n,n^\prime-1}
-\sqrt{n}\delta_{n,n^\prime+1})\Big]$~(also see Refs.~\cite{qhchen2008pra,jren2012prb}).
Based on the systematic perturbation (see the full solution in part C of the Appendix), the steady-state heat current in the bias regime $[(\varepsilon-\omega_0){\gg}\lambda]$ is described as
\begin{eqnarray}~\label{jzq}
J^z_{ss}\approx\Big(\frac{2\lambda}{\omega_0}\Big)^2\omega_0(I_{z,1}+I_{z,2}),
\end{eqnarray}
where these two components are given by
\begin{subequations}
\begin{align}
I_{z,1}=&
\frac{\gamma_\textrm{Q}(\varepsilon+\omega_0)}{2n_{\textrm{Q}}(\varepsilon)+1}\{[1+n_{\textrm{Q}}(\varepsilon+\omega_0)]n_{\textrm{Q}}(\varepsilon)n_{\textrm{R}}(\omega_0)\nonumber\\
&-n_{\textrm{Q}}(\varepsilon+\omega_0)[1+n_{\textrm{Q}}(\varepsilon)][1+n_{\textrm{R}}(\omega_0)]\},~\label{iz1}\\
I_{z,2}=&
\frac{\gamma_\textrm{Q}(\varepsilon-\omega_0)}{2n_{\textrm{Q}}(\varepsilon)+1}\{n_{\textrm{Q}}(\varepsilon-\omega_0)[1+n_{\textrm{Q}}(\varepsilon)]n_{\textrm{R}}(\omega_0)\nonumber\\
&-[1+n_{\textrm{Q}}(\varepsilon-\omega_0)]n_{\textrm{Q}}(\varepsilon)[1+n_{\textrm{R}}(\omega_0)]\}.~\label{iz2}
\end{align}
\end{subequations}
In sharp contrast to $\theta=0$ limit, $I_{z,1}$ is contributed by two competing cyclic fluxes,
 rather not the cotunneling processes.
Specifically, the first term $[1+n_{\textrm{Q}}(\varepsilon+\omega_0)]n_{\textrm{Q}}(\varepsilon)n_{\textrm{R}}(\omega_0)$
describes the loop transition
$|\phi^{\uparrow}_{n+1}{\rangle}{\rightarrow}|\phi^{\downarrow}_n{\rangle}{\rightarrow}|\phi^\uparrow_n{\rangle}{\rightarrow}|\phi^\uparrow_{n+1}{\rangle}$ by directionally transferring the energy $\omega_0$  into the $\textrm{Q}$-th thermal reservoir,
which is also depicted in Fig.~\ref{fig1}(c).
And the second term
$n_{\textrm{Q}}(\varepsilon+\omega_0)[1+n_{\textrm{Q}}(\varepsilon)][1+n_{\textrm{R}}(\omega_0)]$
shows the counter loop transition.
Similarly, $I_{z,2}$ is composed by other two opposite cyclic fluxes, where the first loop path is shown in Fig.~\ref{fig1}(d), and characterizes the joint transport process
$|\phi^{\uparrow}_{n+1}{\rangle}{\rightarrow}|\phi^{\downarrow}_{n+1}{\rangle}{\rightarrow}
|\phi^{\uparrow}_{n}{\rangle}{\rightarrow}|\phi^{\uparrow}_{n+1}{\rangle}$.
Intriguingly, at large temperature bias ${\Delta}T{\approx}2\omega_0$,
i.e., $T_{\textrm{R}}{\approx}2\omega_0$ and $T_{\textrm{Q}}{\approx}0$,
All of cyclic current components in $I_{z,1}$ and $I_{z,2}$ break down, due to negligible excitation in the $\textrm{Q}$-th reservoir
[$n_{\textrm{Q}}(\omega>0){\approx}0$].
This directly results in the suppression of the steady-state heat current, which identifies the emergence of the NDTC.

Therefore, we exploit two distinct microscopic pictures in limiting composite angles with weak qubit-resonator interaction,
i.e., cotunneling transitions at $\theta=0$ and cyclic transitions at $\theta=\pi/2$,
which are generic to unravel microscopic mechanisms of quantum thermal transport,
e.g., in nonequilibrium spin-boson model~\cite{truokola2011prb,cwang2015sr,dzxu2016fop,cwang2017pra,tchen2020pre,crduan2020jpcl},
metal-insulator interfaces~\cite{jren2013prb,jren2013prbr}, and inelastic thermoelectrics~\cite{jhjiang2015prb,jhjiang2017prapp,jclu2020prb,jclu2021prb}.
The analytical expressions of the heat current in Eq.~(\ref{jxss})
and Eq.~(\ref{jzq}) are obtained for the first time in dissipative qubit-resonator hybrid model.
Moreover, though the transport picture with longitudinal qubit-resonator coupling was preliminarily reported in Ref.~\cite{cwang2021cpl},
it was analyzed based on only transition rates.

\subsubsection{Strong qubit-resonator interaction}
\begin{figure}[tbp]
\begin{center}
\includegraphics[scale=0.45]{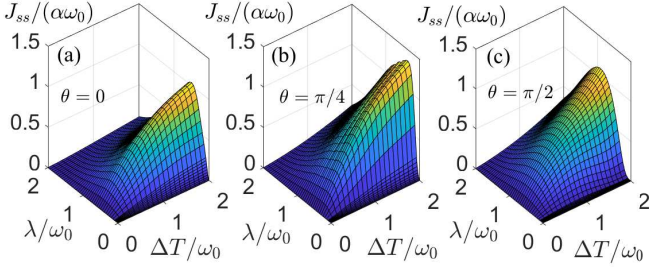}
\end{center}
\caption{(Color online)
Steady-state heat current $J_{ss}/(\alpha\omega_0)$ modulated by temperature bias ${\Delta}T/\omega_0$ and qubit-resonator interaction strength $\lambda/\omega_0$ with composite angle (a) $\theta=0$, (b) $\theta=\pi/4$, and (c) $\theta=\pi/2$.
Other parameters are given by $\varepsilon=1.5\omega_0$, $\alpha=0.001$,  $\omega_c=10\omega_0$,
$T_{\textrm{R}}=\omega_0+{{\Delta}T}/2$, and $T_{\textrm{Q}}=\omega_0-{{\Delta}T}/2$.
}~\label{fig2-theta}
\end{figure}

\begin{figure}[tbp]
\begin{center}
\includegraphics[scale=0.4]{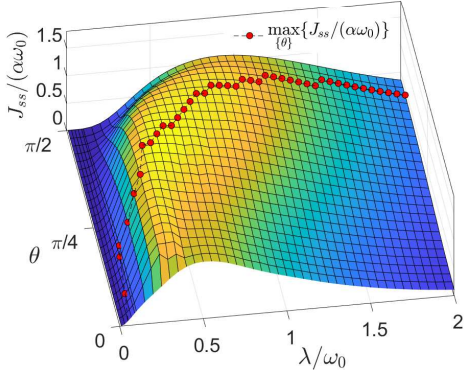}
\end{center}
\caption{(Color online) Steady-state heat current $J_{ss}/(\alpha\omega_0)$ modulated by  qubit-resonator interaction strength $\lambda/\omega_0$ and  composite angle $\theta$.
The dashed line with red circles shows the maximal $J_{ss}/(\alpha\omega_0)$ by searching $\theta{\in}[0,\pi/2]$ with given $\lambda/\omega_0$.
Other parameters are given by $\varepsilon=1.5\omega_0$, $\alpha=0.001$,  $\omega_c=10\omega_0$,
$T_{\textrm{R}}=2\omega_0$, and $T_{\textrm{Q}}=0$.
}~\label{fig2-dT}
\end{figure}

Next, we investigate steady-state heat current $J_{ss}/(\alpha\omega_0)$ beyond weak qubit-resonator coupling with typical composite angles in Fig.~\ref{fig2-theta}.
As the qubit shows transverse interaction with the optical resonator, i.e., $\theta=0$ in Fig.~\ref{fig2-theta}(a),
the heat current exhibits nonmonotonic behavior by increasing qubit-resonator interaction strength,
under finite temperature bias (e.g., ${\Delta}T/\omega_0=1$).
Due to the effect of counter-rotating-terms, the eigenstates of the quantum Rabi model beyond Eqs.~(\ref{vecxp}-\ref{vecxm})
will introduce additional energy exchange transitions,
which may effectively enhance heat current in the regime $\lambda/\omega_0{\lesssim}0.5$.
While in strong qubit-resonator coupling limit (e.g., $\lambda/\omega_0>1.5$),
the eigenstates become nearly degenerate, i.e.,
$|\phi^{\pm}_n{\rangle}{\approx}\exp[\pm\lambda(\hat{a}-\hat{a}^\dag)/\omega_0][(\hat{a}^\dag)^n/\sqrt{n!}]|0{\rangle}{\otimes}|\pm{\rangle}$
with $\hat{\sigma}_x|\pm{\rangle}=\pm|\pm{\rangle}$,
which significantly prohibits energy exchange between the hybrid system and the $\textrm{Q}$-th reservoir
(${\langle}\phi^{+}_n|\hat{\sigma}_x|\phi^{-}_n{\rangle}{\approx}0$).
Then, by tuning on the composite angle, e.g., $\theta=\pi/4$ and $\pi/2$, it is found that
the profiles of heat currents modulated by  ${\Delta}T/\omega_0$
and $\lambda/\omega_0$ in Figs.~\ref{fig2-theta}(b-c) are generally similar with limiting angle case $\theta=0$.
In particular for the limiting case $\theta=\pi/2$, the transition coefficient in Eqs.~(\ref{ratep}-\ref{ratem}) is expressed as~\cite{qhchen2008pra,jren2012prb}
\begin{eqnarray}~\label{szoverlap}
{\langle}\phi^{\uparrow}_{n}|\hat{\sigma}_x|\phi^{\downarrow}_{n^{\prime}}{\rangle}&=&
{(-1)}^n\exp{(-2\lambda^2/\omega^2_0)}\sqrt{n!n^{\prime}!}\nonumber\\
&&{\times}\sum^{\min[n,n^\prime]}_{l=0}\frac{(-1)^{l}\sqrt{(2\lambda/\omega_0)^{n+n^\prime-2l}}}
{(n-l)!(n^\prime-l)!l!},
\end{eqnarray}
which induces higher-order transitions between
$|\phi^{\uparrow(\downarrow)}_{n}{\rangle}$ and $|\phi^{\downarrow(\uparrow)}_{n^\prime}{\rangle}$ with $|n-n^\prime|{\ge}2$, besides the lowest-order transport processes, i.e., cyclic exchange in Figs.~\ref{fig1}(c-d).
These additional transitions are robust even at large temperature bias,
which mainly result in comparatively large heat current and disappearance of the NDTC.
While
at finite temperature bias (e.g., ${\Delta}T/\omega_0=1$),
the initial enhancement of $J_{ss}/(\alpha\omega_0)$ by increasing $\lambda/\omega_0$ stems from additional transport processes.
It may be quantified by
${\langle}\phi^{\uparrow}_{n}|\hat{\sigma}_x|\phi^{\downarrow}_{n^{\prime}}{\rangle}
{\approx}(-1)^n\Big\{[1-(n+1/2)(\frac{2\lambda}{\omega_0})^2]\delta_{n,n^\prime}
+(\frac{2\lambda}{\omega_0})(\sqrt{n+1}\delta_{n,n^\prime-1}-\sqrt{n}\delta_{n,n^\prime+1})
+\frac{1}{2}(\frac{2\lambda}{\omega_0})^2[\sqrt{n(n-1)}\delta_{n,n^\prime+2}
+\sqrt{(n+1)(n+2)}\delta_{n,n^\prime-2}]\Big\}$,
where last two terms  will enhance the current by forming efficient transition paths.
However, the final decrease of heat current in strong qubit-resonator coupling limit is mainly attributed to the fact  the transition coefficient ${\langle}\phi^{\uparrow}_{n}|\hat{\sigma}_x|\phi^{\downarrow}_{n^{\prime}}{\rangle}$ in Eq.~(\ref{szoverlap}) and the corresponding transition rates are dramatically weakened.
Consequently, the energy exchange processes are strongly blocked.

Moreover, we analyze the interplay between the composite angle and qubit-resonator coupling strength on
$\max_{\{\theta\}}\{J_{ss}/(\alpha\omega_0)\}$ in Fig.~\ref{fig2-dT}.
It is intriguing to find that with increase of the qubit-resonator interaction,
the composite angle dominating the maximal heat current is gradually modified from $0$ to $\pi/2$,
 corresponding to the transverse and longitudinal qubit-resonator couplings, respectively.
Hence, we conclude that at strong $\lambda/\omega_0$, the longitudinal coupling type is preferred to enhance the steady-state heat current.



\subsection{Two-photon correlation function}
Two-photon correlation function describes the correlation between two temporally separated photon signals  from one light source,
which is pioneered by R. J. Glauber to unveiling the optical coherence of quantum theory~\cite{rjglauber1963pr}.
Later, it has been extended to investigate superradiant spontaneous emission~\cite{jheberly1970pra,rbonifacio1971pra,rbonifacio1971pra2,dmeiser2010pra,AAuffeves2011njp,EMascarenhas2013pra},
strongly interacting photons~\cite{aimamoglu1997prl,prabl2011prl,droy2017rmp},
and photon (phonon) bundle emission~\cite{qbin2020prl,qbin2021prl}.
Alternatively, Ridolfo \emph{et al.}~\cite{aridolfo2012prl,aridolfo2013prl,rstassi2013prl,lgarziano2017acs} proposed a modified definition of two-photon correlation function within the dressed framework.
Hence, it can also be safely included to study photon statistics in the present model in Eq.~(\ref{htot})
with strong qubit-resonator interaction.
Specifically, two-photon correlation function at steady state is defined as
\begin{eqnarray}
G^{(2)}_\theta(\tau)=\lim_{t{\rightarrow}\infty}\frac{{\langle}\hat{X}^+_{\theta}(t)\hat{X}^+_{\theta}(t+\tau)\hat{X}^-_{\theta}(t+\tau)
\hat{X}^-_{\theta}(t){\rangle}}{{\langle}\hat{X}^+_{\theta}(t)\hat{X}^-_{\theta}(t){\rangle}^2},
\end{eqnarray}
where the measurement operator is $\hat{X}^-_\theta=-i\sum_{k>j}\Delta_{kj}X_{jk}|\phi_j{\rangle}{\langle}\phi_k|$
and $\hat{X}^+_\theta=(\hat{X}^-_\theta)^\dag$,
with $X_{jk}={\langle}\phi_j|(\hat{a}^\dag+\hat{a})|\phi_k{\rangle}$ the energy gap $\Delta_{jk}=E_j-E_k$, and the eigensolution
$\hat{H}^\theta_{\textrm{S}}|\phi_k{\rangle}=E_k|\phi_k{\rangle}$.
Here, we focus on the effect of the composite qubit-resonant interaction on zero-time two-photon correlation function
$G^{(2)}_\theta(0)$.

\begin{figure}[tbp]
\begin{center}
\includegraphics[scale=0.4]{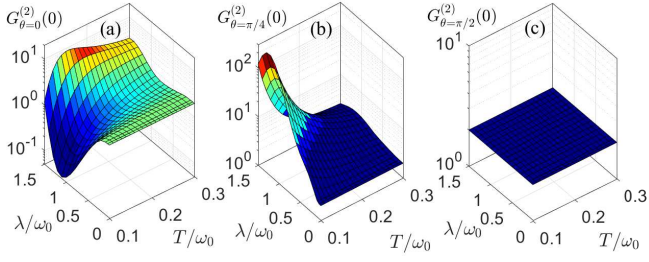}
\end{center}
\caption{(Color online) Two-photon correlation function as functions of  qubit-resonator interaction strength $\lambda/\omega_0$ and
temperature $T_{\textrm{R}}=T_{\textrm{Q}}=T$, with typical composite angle (a) $\theta=0$, (b) $\theta=\pi/4$, and (c) $\theta=\pi/2$.
Other parameters are given by $\varepsilon=1.5\omega_0$, $\alpha=0.001$,  and $\omega_c=10\omega_0$.
}~\label{fig3-theta}
\end{figure}

\begin{figure}[tbp]
\begin{center}
\includegraphics[scale=0.5]{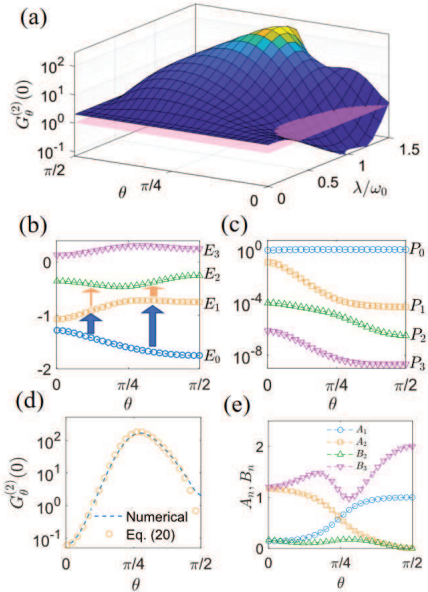}
\end{center}
\caption{(Color online) (a) Two-photon correlation function $G^{(2)}_\theta(0)$  modulated by  qubit-resonator interaction strength $\lambda/\omega_0$ and  composite angle $\theta$ in low temperature regime.
(b) Comparison of $G^{(2)}_\theta(0)$ via numerical calculation with that based on Eq.~(*),
(c) four lowest eigenenergies, (d) four lowest steady-state populations, and (e) coefficients $A_{n},~B_{n}$ as a function of composite angle $\theta$ at $\lambda=\omega_0$.
Other parameters are given by $\varepsilon=1.5\omega_0$, $\alpha=0.001$,  $\omega_c=10\omega_0$, and
$T_{\textrm{R}}=T_{\textrm{Q}}=0.1\omega_0$.
}~\label{fig4}
\end{figure}

We first investigate two-photon correlation function by tuning both the qubit-resonator coupling strength and bath temperature with different composite angle in Fig.~\ref{fig3-theta}.
For $\theta=0$, it is found that
an intriguing antibunching behavior can be found with strong $\lambda$, which demonstrates the seminal two-photon blockade~\cite{aridolfo2012prl}.
Then, by tuning on composite angle, e.g., $\theta=\pi/4$, a giant photon bunching behavior is exhibited at low temperature.
If we further increase the composite angle to $\theta=\pi/2$, it is shown that $G^{(2)}_{\theta=0}(0){\approx}2$
regardless of $\lambda$ and $T$ ($T_{\textrm{R}}=T_{\textrm{Q}}=T$),
due to the fully thermalization of the longitudinally coupled qubit-resonator system
\begin{eqnarray}~\label{rhoss}
\rho_{\textrm{S}}(\infty)&=&\frac{\sinh[\omega_0/(2T)]}{\cosh[\varepsilon/(2T)]}
\sum_{n}e^{-\frac{(n+1/2)\omega_0}{T}}\nonumber\\
&&{\times}[e^{-\frac{\varepsilon}{2T}}|\phi^\uparrow_n{\rangle}
{\langle}\phi^\uparrow_n|+e^{\frac{\varepsilon}{2T}}]|\phi^\downarrow_n{\rangle}
{\langle}\phi^\downarrow_n|].
\end{eqnarray}
Moreover, we plot Fig.~\ref{fig4}(a) to see the influence of the composite angle in two-photon correlation function.
It is found that at strong $\lambda$ (e.g., $\lambda/\omega_0{\approx}1$), an antibunching-to-bunching transition is clearly exhibited by increasing $\theta$.
Therefore, we conclude that the modulation of the composite angle is quite important to exhibit the nonclassical photon statistics,
which may provide physical guidance to measure photon correlation in circuit QED.

Next, we analyze the mechanism of antibunching-to-bunching  transition modulated by the composite angle.
At low temperature (e.g., $T=0.1\omega_0$), the finite spacing distribution of energy levels [see Fig.~\ref{fig4}(b)] results in
$P_0{\gg}P_1{\gg}P_2{\gg}P_3$, exhibited in Fig.~\ref{fig4}(c).
Hence, one-photon and two-photon terms are approximated as
${\langle}\hat{X}^+_\theta\hat{X}^-_\theta{\rangle}{\approx}P_1A_1$
and
${\langle}(\hat{X}^+_\theta)^2(\hat{X}^-_\theta)^2{\rangle}{\approx}P_2B_2$,
where the coefficients are
$A_n=\sum_{l<k}(\Delta_{kl}X_{kl})^2$
and
$B_n=\sum_{p<l<k}(\Delta_{kl}\Delta_{lp}X_{kl}X_{lp})^2$.
Consequently, the two-photon correlation function is expressed in a concise way
\begin{eqnarray}~\label{g2simp}
G^{(2)}_\theta(0){\approx}\frac{P_2B_2}{(P_1A_1)^2},
\end{eqnarray}
which shows agreement with the numerical result in Fig.~\ref{fig4}(d) in a wide regime of $\theta$ at strong qubit-resonator coupling ($\lambda/\omega_0=1$).
Particularly in the anti-bunching regime (e.g., $0<\theta{{\lesssim}}\pi/10$),
the large energy gap $(E_2-E_1)$ suppresses the ratio $P_2/P^2_1$.
Moreover, the coefficients $A_1$ and $B_2$ are nearly flat, as shown in Fig.~\ref{fig4}(e).
Hence, the transition $|\phi_1{\rangle}{\rightarrow}|\phi_2{\rangle}$ is strongly blocked,
leading to the antibunching behavior of photons.
While in the bunching regime (e.g., $\pi/10{\lesssim}\theta{\le}\pi/4$),
though $A_1$ is strengthened by increasing $\theta$,
the reduction of energy gap between $E_2$ and $E_1$ dramatically enhances steady-state population $P_2$
and successive two-photon excitation process $|\phi_0{\rangle}{\rightarrow}|\phi_1{\rangle}{\rightarrow}|\phi_2{\rangle}$,
resulting in the bunching behavior of photons, as shown in Fig.~\ref{fig4}(b).




\section{Conclusion}
In summary, we investigate the effect of composite qubit-resonator interaction on quantum thermal transport and zero-delay-delay two-photon correlation function at steady state.
We apply the quantum dressed master equation to properly treat strong qubit-resonator interaction with arbitrary composite angle.
For heat transport at weak qubit-resonator coupling, it is found that the heat current with transverse qubit-resonator coupling shows monotonic behavior by increasing bath temperature bias.
It is dominated by the cotunneling process, which is quantified by Eqs.~(\ref{ix1}-\ref{ix2}).
While the current is gradually changed to exhibit nonmonotonic feature by tuning on the composite angle, signifying the emergence of the NDTC effect, completely characterized as the cyclic energy exchange processes, and described by Eqs.~(\ref{iz1}-\ref{iz2}).
Hence, we unravel two crucial microscopic processes for quantum thermal transport.
In strong qubit-resonator interaction regime,
the heat current with $\theta=0$ exhibits nonmonotonic behavior by increasing qubit-resonator interaction strength.
The initial enhancement of $J_{ss}$ stems from additional energy transitions due to counter-rotating-terms,
whereas the final suppression of $J_{ss}$ is attributed to the nearly degeneracy of eigenstates,
which prohibits energy exchange between the qubit and $\textrm{Q}$-th thermal reservoir.
Moreover, it is intriguing to find that the optimal composite angle, which corresponds to the maximal heat current, switches directionally from $\theta=0$ (transverse) to $\theta=\pi/2$ (longitudinal) by modulating qubit-resonator interaction from weak to strong couplings.
Hence, the longitudinal coupling is preferred to enhance the steady-state heat current in strong qubit-resonator interaction regime.

We also investigate the steady-state two-photon correlation function by modulating the composite angle.
At $\theta=0$, the pronounced anti-bunching feature is exhibited in the regime of low temperature and strong qubit-resonator interaction.
Then by tuning on $\theta$, we find the giant bunching signal instead.
By further increasing $\theta$ (e.g., $\theta=\pi/2$), two-photon correlation function is globally around $2$,
due to special distribution of the density operator of the qubit-resonator hybrid system in Eq.~(\ref{rhoss}).
Moreover, We also present the mechanism with approximate expression in Eq.~(\ref{g2simp}) to explain
this antibunching-to-bunching transition.
The antibunching and bunching behaviors of photons
are modulated by the large and reduced energy gap between  the first and second excited eigenstates, respectively.
We hope that our results affected by composite qubit-resonator interaction may deepen the understanding of quantum thermal transport
and two-photon statistics in dissipative QED systems.

\section*{ACKNOWLEDGEMENTS}

Z.C., H.C, Z.C, and C.W. are supported by the
National Natural Science Foundation
of China under Grant No. 11704093
and the Opening Project of Shanghai Key Laboratory of Special Artificial Microstructure Materials and Technology.
J.R. acknowledges the support
by the National Natural Science Foundation of China
(No. 11935010 and No. 11775159), Natural Science
Foundation of Shanghai (No. 18ZR1442800 and No. 18JC1410900).

\section*{Appendix}
\subsection{General expression of heat current}
From the quantum dressed master equation Eq.~(\ref{dme}), the population dynamics is described as
\begin{eqnarray}~\label{rhome}
\frac{d}{dt}\rho_{nn}(t)=\sum_{n,m,\mu}[\Gamma^-_\mu(E_{n,m})\rho_{mm}(t)-\Gamma^+_\mu(E_{n,m})\rho_{nn}(t)],
\end{eqnarray}
with the transition rates $\Gamma^{\pm}_\mu(E_{n,m})$ specified by Eqs.~(\ref{ratep}-\ref{ratem}),
and the population $\rho_{nn}(t)={\langle}\phi_n|\hat{\rho}_{\textrm{S}}(t)|\phi_n{\rangle}$.
Hence, the expression of steady-state heat current into the $\mu$-th thermal bath is expressed as
\begin{eqnarray}
J_{\mu}=\sum_{E_n>E_{n^\prime}}E_{n,n^\prime}[\Gamma^-_{\mu}(E_{n,n^\prime})P_{nn}-\Gamma^+_{\mu}(E_{n,n^\prime})P_{n^{\prime}n^{\prime}}].
\end{eqnarray}
Then, we try to approximately investigate the steady state populations $P_{nn}={\langle}\phi_n|\hat{\rho}_{\textrm{S}}(t\rightarrow\infty)|\phi_n{\rangle}$ at weak qubit-resonator coupling limit via the perturbation method.
As we reexpress the populations in the vector form $|\rho_{s}{\rangle}$,
the steady-state solution based on Eq.~(\ref{rhome}) becomes $\mathcal{L}|\rho_{s}{\rangle}=0$.
At weak qubit-resonator coupling, we expand
$|\rho_{\textrm{S}}{\rangle}{\approx}|\rho^{(0)}_{\textrm{S}}{\rangle}+(\lambda/\omega_0)^2|\rho^{(1)}_\textrm{S}{\rangle}$
and $\mathcal{L}{\approx}\mathcal{L}^{(0)}+(\lambda/\omega_0)^2\mathcal{L}^{(1)}$
up to $(\lambda/\omega_0)^2$.
Hence, the general solution is given by
\begin{subequations}
\begin{align}
\mathcal{L}^{(0)}|\rho^{(0)}_{\textrm{S}}{\rangle}=&0,~\label{l0}\\
\mathcal{L}^{(0)}|\rho^{(1)}_{\textrm{S}}{\rangle}+\mathcal{L}^{(1)}|\rho^{(0)}_{\textrm{S}}{\rangle}=&0~\label{l1}.
\end{align}
\end{subequations}

\subsection{$\theta=0$ case}
At $\theta=0$, the model in the weak qubit-resonator coupling regime becomes the dissipative Jaynes-Cummings model.
We consider the off-resonant regime $[(\varepsilon-\omega_0){\gg}\lambda]$.
Under the eigenbasis $|\phi^{\pm}_{n}{\rangle}$ [Eqs.~(\ref{vecxp}-\ref{vecxm})] the transition coefficients involved with the $\textrm{Q}$-th bath  are specified as
\begin{subequations}
\begin{align}
{\langle}\phi^+_{n+1}|\hat{\sigma}_x|\phi^+_{n}{\rangle}=&\sin\frac{\varphi_{n}}{2}\cos\frac{\varphi_{n+1}}{2},\\
{\langle}\phi^+_{n+1}|\hat{\sigma}_x|\phi^-_{n}{\rangle}=&\cos\frac{\varphi_{n}}{2}\cos\frac{\varphi_{n+1}}{2},\\
{\langle}\phi^-_{n+1}|\hat{\sigma}_x|\phi^+_{n}{\rangle}=&-\sin\frac{\varphi_{n}}{2}\sin\frac{\varphi_{n+1}}{2},\\
{\langle}\phi^-_{n+1}|\hat{\sigma}_x|\phi^-_{n}{\rangle}=&-\cos\frac{\varphi_{n}}{2}\sin\frac{\varphi_{n+1}}{2},
\end{align}
\end{subequations}
with $\tan\varphi_n=2\lambda\sqrt{n+1}/\Delta$ and $\Delta=\varepsilon-\omega_0$.
Here, we further approximately treat the eigenvalues as $E_{n,+}{\approx}\omega_0(n+1/2)+\varepsilon/2$
and $E_{n,-}{\approx}\omega_0(n+3/2)-\varepsilon/2$.
And the corresponding eigenvectors are simplified to
$|\phi^+_n{\rangle}{\approx}|n{\rangle}{\otimes}|\uparrow{\rangle}$
and
$|\phi^-_n{\rangle}{\approx}|n+1{\rangle}{\otimes}|\downarrow{\rangle}$.
Hence, the transition rates upper to the order $(\lambda/\Delta)^2$ are given by
\begin{subequations}
\begin{align}
\Gamma^{\pm}_{\textrm{Q}}(E^{n+1,+}_{n,+})\approx&\gamma_{\textrm{Q}}(\pm\omega_0)n_{\textrm{Q}}(\pm\omega_0)\Big(\frac{\lambda}{\Delta}\Big)^2(n+1),\\
\Gamma^\pm_{\textrm{Q}}(E^{n+1,+}_{n,-})\approx&\gamma_{\textrm{Q}}(\pm\varepsilon)n_{\textrm{Q}}(\pm\varepsilon)
\Big[1-\Big(\frac{\lambda}{\Delta}\Big)^2(2n+3)\Big],\\
\Gamma^{\pm}_{\textrm{Q}}(E^{n+1,-}_{n,+})\approx&0,\\
\Gamma^{\pm}_{\textrm{Q}}(E^{n+1,-}_{n,-})\approx&\gamma_{\textrm{Q}}(\pm\omega_0)n_{\textrm{Q}}(\pm\omega_0)\Big(\frac{\lambda}{\Delta}\Big)^2(n+2),
\end{align}
\end{subequations}
with $E^{n,\sigma}_{n^\prime,\sigma^\prime}=E_{n,\sigma}-E_{n^\prime,\sigma^\prime}$.
Similarly, the rates related with the $\textrm{R}$-th bath are approximated by
\begin{subequations}
\begin{align}
\Gamma^{\pm}_{\textrm{R}}(E^{n+1,+}_{n,+})\approx&\gamma_{\textrm{R}}(\pm\omega_0)n_{\textrm{R}}(\pm\omega_0)(n+1)
\Big[1+\Big(\frac{\lambda}{\Delta}\Big)^2\Big],\\
\Gamma^\pm_{\textrm{R}}(E^{n+1,+}_{n,-})\approx&\gamma_{\textrm{Q}}(\pm\varepsilon)n_{\textrm{Q}}(\pm\varepsilon)
\Big(\frac{\lambda}{\Delta}\Big)^2,\\
\Gamma^{\pm}_{\textrm{R}}(E^{n,-}_{n-1,+})\approx&0,\\
\Gamma^{\pm}_{\textrm{R}}(E^{n,-}_{n-1,-})\approx&\gamma_{\textrm{Q}}(\pm\omega_0)n_{\textrm{Q}}(\pm\omega_0)(n+2)
\Big[1-\Big(\frac{\lambda}{\Delta}\Big)^2\Big].
\end{align}
\end{subequations}
From Eq.~(\ref{l1}) we have the relation at steady state
\begin{eqnarray}
&&\sum_n[\Gamma^-_{\textrm{Q}}(E^{n,+}_{n-1,-})P_{n,+}-\Gamma^+_{\textrm{Q}}(E^{n,+}_{n-1,-})P_{n-1,-}]\nonumber\\
&=&\sum_{n}[\Gamma^+_{\textrm{R}}(E^{n,+}_{n-1,-})P_{n-1,-}-\Gamma^-_{\textrm{R}}(E^{n,+}_{n-1,-})P_{n,+}]\nonumber\\
&&+\sum_{n}[\Gamma^+_{\textrm{Q}}(E^{n,+}_{n-1,+})P_{n-1,+}-\Gamma^-_{\textrm{Q}}(E^{n,+}_{n-1,+})P_{n,+}]\nonumber\\
&&+\sum_{n}[\Gamma^-_{\textrm{Q}}(E^{n+1,+}_{n,+})P_{n+1,+}-\Gamma^+_{\textrm{Q}}(E^{n+1,+}_{n,+})P_{n,+}].\nonumber
\end{eqnarray}
Hence, the zeroth order populations are given by
\begin{subequations}
\begin{align}
P^{(0)}_{n,+}\approx&\frac{(1-e^{-\beta_\textrm{R}\omega_0})}{e^{\beta_\textrm{Q}\varepsilon}+1}e^{-{n}\beta_\textrm{R}\omega_0},\\
P^{(0)}_{n,-}\approx&\frac{e^{\beta_\textrm{Q}\varepsilon}(1-e^{-\beta_\textrm{R}\omega_0})}{e^{\beta_\textrm{Q}\varepsilon}+1}e^{-{(n+1)}\beta_\textrm{R}\omega_0},
\end{align}
\end{subequations}
with $\beta_u=1/(T_u)~(u=\textrm{R}, \textrm{Q})$.
Consequently, we obtain the expression of heat current as
\begin{eqnarray}
J^{x}_{\textrm{Q}}&=&\Big(\frac{\lambda}{\varepsilon-\omega_0}\Big)^2(\omega_0I_{x,1}+{\varepsilon}I_{x,2}),
\end{eqnarray}
where two components are specified as
\begin{subequations}
\begin{align}
I_{x,1}=&\gamma_{\textrm{Q}}(\omega_0)\{n_{\textrm{R}}(\omega_0)[1+n_{\textrm{Q}}(\omega_0)]-[1+n_{\textrm{R}}(\omega_0)]n_{\textrm{Q}}(\omega_0)\},\\
I_{x,2}=&\frac{\gamma_{\textrm{R}}(\varepsilon)}{2n_{\textrm{Q}}(\varepsilon)+1}
\{n_{\textrm{R}}(\varepsilon)[1+n_{\textrm{Q}}(\varepsilon)]
-{[1+n_{\textrm{R}}(\varepsilon)]}n_{\textrm{Q}}(\varepsilon)\},
\end{align}
\end{subequations}

\subsection{$\theta=\pi/2$ case}

At $\theta=\pi/2$, Based on the eigenbasis $|\phi^{\uparrow(\downarrow)}_{n}{\rangle}$ at Eqs.~(\ref{veczp}-\ref{veczm}) the transition coefficient involved with the $\textrm{Q}$-th bath is simplified to
\begin{eqnarray}
{\langle}\phi^{\sigma}_{n}|\hat{\sigma}_x|\phi^{\overline{\sigma}}_{n^{\prime}}{\rangle}
&{\approx}&(-1)^n\Big[\delta_{n,n^\prime}+\frac{2\lambda}{\omega_0}\sqrt{n+1}\delta_{n,n^\prime-1}\nonumber\\
&&-\frac{2\lambda}{\omega_0}\sqrt{n}\delta_{n,n^\prime+1}\Big].
\end{eqnarray}
Accordingly, the transition rates defined by Eqs.~(\ref{ratep}-\ref{ratem}) are approximated by
\begin{subequations}
\begin{align}
\Gamma^{\pm}_{\textrm{Q}}(E^{n,\uparrow}_{n^\prime,\downarrow})\approx&\delta_{n,n^\prime}\kappa^{\pm}_{\textrm{Q}}(\varepsilon)
+\delta_{n,n^\prime-1}n^\prime\Big(\frac{2\lambda}{\omega_0}\Big)^2\kappa^{\pm}_{\textrm{Q}}(\varepsilon-\omega_0)\\
&+\delta_{n,n^\prime+1}n\Big(\frac{2\lambda}{\omega_0}\Big)^2\kappa^{\pm}_{\textrm{Q}}(\varepsilon+\omega_0),\nonumber\\
\Gamma^{\pm}_{\textrm{Q}}(E^{n,\downarrow}_{n^\prime,\uparrow})\approx&\delta_{n,n^\prime+1}n^\prime\Big(\frac{2\lambda}{\omega_0}\Big)^2\kappa^{\pm}_{\textrm{Q}}(\omega_0-\varepsilon),
\end{align}
\end{subequations}
with
$\kappa^{+}_{\textrm{Q}}(\omega)=\gamma_{\textrm{Q}}(\omega)n_{\textrm{Q}}(\omega)$
and
$\kappa^{-}_{\textrm{Q}}(\omega)=\gamma_{\textrm{Q}}(\omega)[1+n_{\textrm{Q}}(\omega)]$.
Similarly,  the transition rates assisted by the $\textrm{R}$-th thermal bath are given by
$\Gamma^{+}_{\textrm{R}}(E^{n,\sigma}_{n-1,\sigma})=\gamma_{\textrm{R}}(\omega_0)n_{\textrm{R}}(\omega_0)n$
and
$\Gamma^{-}_{\textrm{R}}(E^{n,\sigma}_{n-1,\sigma})=\gamma_{\textrm{R}}(\omega_0)[1+n_{\textrm{R}}(\omega_0)]n$.
Hence, the zeroth order of populations based on Eq.~(\ref{l0}) can be directly obtained as
\begin{subequations}
\begin{align}
P^{(0)}_{n,\uparrow}\approx&\frac{(1-e^{-\beta_\textrm{R}\omega_0})}{e^{\beta_\textrm{Q}\varepsilon}+1}e^{-{n}\beta_\textrm{R}\omega_0},\\
P^{(0)}_{n,\downarrow}\approx&\frac{e^{\beta_\textrm{Q}\varepsilon}(1-e^{-\beta_\textrm{R}\omega_0})}{e^{\beta_\textrm{Q}\varepsilon}+1}e^{-{n}\beta_\textrm{R}\omega_0}.
\end{align}
\end{subequations}
Moreover, from Eq.~(\ref{l1}) it is known that
\begin{eqnarray}
&&\Big(\frac{2\lambda}{\omega_0}\Big)^2\sum_n[\Gamma^-_\textrm{Q}(E^{n,\uparrow}_{n,\downarrow})P^{(1)}_{n,\uparrow}
-\Gamma^+_\textrm{Q}(E^{n,\uparrow}_{n,\downarrow})P^{(1)}_{n\downarrow}]
\approx\nonumber\\
&&\sum_n[-\Gamma^-_\textrm{Q}(E^{n,\uparrow}_{n-1,\downarrow})P^{(0)}_{n,\uparrow}
+\Gamma^+_\textrm{Q}(E^{n,\uparrow}_{n-1,\downarrow})P^{(0)}_{n-1,\downarrow}\nonumber\\
&&-\Gamma^-_\textrm{Q}(E^{n,\uparrow}_{n+1,\downarrow})P^{(0)}_{n,\uparrow}
+\Gamma^+_\textrm{Q}(E^{n,\uparrow}_{n+1,\downarrow})P^{(0)}_{n+1,\downarrow}\nonumber\\
&&-\Gamma^+_\textrm{Q}(E^{n+1,\downarrow}_{n,\uparrow})P^{(0)}_{n+1,\downarrow}
+\Gamma^-_\textrm{Q}(E^{n+1,\downarrow}_{n,\uparrow})P^{(0)}_{n,\uparrow}].\nonumber
\end{eqnarray}
Finally,
Then, the current is contributed by three components
\begin{eqnarray}
J^z_{\textrm{Q}}\approx\Big(\frac{2\lambda}{\omega_0}\Big)^2\omega_0(I_{z,1}+I_{z,2}+I_{z,3}),
\end{eqnarray}
where
\begin{subequations}
\begin{align}
I_{z,1}=&\theta(\varepsilon+\omega_0)\gamma_\textrm{Q}(\varepsilon+\omega_0)
\frac{1}{2n_{\textrm{Q}}(\varepsilon)+1}\nonumber\\
&{\times}[(1+n_{\textrm{Q}}(\varepsilon+\omega_0))n_{\textrm{Q}}(\varepsilon)n_{\textrm{R}}(\omega_0)\nonumber\\
&-n_{\textrm{Q}}(\varepsilon+\omega_0)(1+n_{\textrm{Q}}(\varepsilon))(1+n_{\textrm{R}}(\omega_0))],\\
I_{z,2}=&\theta(\varepsilon-\omega_0)\gamma_\textrm{Q}(\varepsilon-\omega_0)
\frac{1}{2n_{\textrm{Q}}(\varepsilon)+1}\nonumber\\
&{\times}[n_{\textrm{Q}}(\varepsilon-\omega_0)(1+n_{\textrm{Q}}(\varepsilon))n_{\textrm{R}}(\omega_0)\nonumber\\
&-(1+n_{\textrm{Q}}(\varepsilon-\omega_0))n_{\textrm{Q}}(\varepsilon)(1+n_{\textrm{R}}(\omega_0))],\\
I_{z,3}=&\theta(\omega_0-\varepsilon)\gamma_\textrm{Q}(\omega_0-\varepsilon)
\frac{1}{2n_{\textrm{Q}}(\varepsilon)+1}\nonumber\\
&{\times}[(1+n_{\textrm{Q}}(\omega_0-\varepsilon))(1+n_{\textrm{Q}}(\varepsilon))n_{\textrm{R}}(\omega_0)\nonumber\\
&-n_{\textrm{Q}}(\omega_0-\varepsilon)n_{\textrm{Q}}(\varepsilon)(1+n_{\textrm{R}}(\omega_0))].
\end{align}
\end{subequations}
with the Heviside function $\theta(x)=1$ for $x{\ge}0$, and $\theta(x)=0$ for $x{<}0$.

\end{document}